\documentclass[a4paper,11pt]{article}
\usepackage[utf8]{inputenc}
\usepackage{jcappub} 
\usepackage{lineno}
\usepackage{slashed}
\usepackage{xcolor}
\usepackage{comment}

\definecolor{darkgreen}{rgb}{0.0, 0.5, 0.0}

\title{Neutrino transition dipole moments in light of the KM3NeT event}

\author[a,b]{Alejandro Muñoz-Ovalle}
\affiliation[a]{Departament de Física Teòrica, Universitat de València, 46100 Burjassot, Spain}

\affiliation[b]{Instituto de Física Corpuscular (CSIC-Universitat de València),
Parc Científic UV, C/Catedrático José Beltrán, 2, E-46980 Paterna, Spain}

\author[c]{ and Stefan Vogl}
\affiliation[c]{Institute of Physics, University of Freiburg, Hermann-Herder-Strasse 3, 79102 Freiburg, Germany}

\emailAdd{almuo@ific.uv.es}
\emailAdd{stefan.vogl@physik.uni-freiburg.de}

\abstract{
In 2023, KM3NeT observed the highest energy neutrino seen to date. The most probable estimate places the energy of the neutrino at an extreme value of 220 PeV, and energies up to 2.6 EeV are contained in the $90\%$ probability interval. 
Even though the source of the signal cannot be ascertained with the available data, this observation already opens unique possibilities to test new physics.
Here, we study the implications for the transition magnetic dipole portal connecting active and sterile neutrinos. We remain agnostic regarding the origin of the neutrino and focus on local effects. Concretely, we study the impact of efficient active--sterile transitions on the energy degradation experienced by the neutrino along its path to the detector at Earth. We find that the ultra-high-energy neutrino event does not add relevant information for sterile neutrinos with masses less than a TeV. However, the observation becomes important at higher masses and can provide the leading constraint in that regime.  
}

\begin{document}
\maketitle
\flushbottom

\section{INTRODUCTION}

The KM3NeT Collaboration has recently reported the observation of a tracklike neutrino event, labeled KM3NeT-230213A, with an estimated energy well above all previously
detected neutrino events~\cite{KM3NeT:2025npi}. With a likely energy exceeding $100~\mathrm{PeV}$, this is by far the most energetic neutrino detected to date, and its unexpected observation has sparked significant interest across both astrophysics and particle physics. On the astrophysical side, such an event may shed light on the nature of the most powerful cosmic accelerators; on the particle physics side, it provides a rare probe of new phenomena at energy scales beyond the reach of terrestrial experiments.  

Several interpretations have been put forward for the origin of this event. Perhaps the simplest explanation is that KM3-230213A could be part of the diffuse flux of cosmogenic (Greisen-Zatsepin-Kuzmin) neutrinos, generated when ultrahigh-energy cosmic rays interact with the cosmic microwave or infrared backgrounds via photopion production. A flux of such neutrinos is expected in the PeV to EeV range and remains compatible with current experimental constraints~\cite{KoteraOlinto2011,Aartsen2018,KM3NeT_ARCA_Astronomy_2024}.  Alternatively, the event might originate from a concrete source.
Dedicated searches for astrophysical counterparts within the reconstructed angular uncertainty revealed 17 candidate blazars, three of which showed multiwavelength activity (radio, x-ray, and
gamma-ray flares) temporally coincident with the neutrino detection, though no definitive association could be established~\cite{km3netcollaboration2025blazar}. Another possibility is that the event originated from a transient flare shorter than one year, which would reconcile the KM3NeT detection with IceCube’s nonobservation more easily than other interpretations~\cite{neronov2025km3230213aultrahighenergyneutrino}. 

Explanations invoking new physics have also been explored. These include scenarios such as neutrinos from primordial black holes~\cite{Boccia:2025hpm,Klipfel:2025jql,Anchordoqui:2025xug} (see, however, ~\cite{Airoldi:2025opo}), decaying dark matter~\cite{Kohri:2025bsn,Borah:2025igh,Murase:2025uwv,Barman:2025hoz}, or exotic contributions to neutrino production in cosmic-ray collisions~\cite{Alves:2025xul}. 
Even if the astrophysical origin remains uncertain, the extreme energy of KM3-230213A makes it a powerful probe of new physics affecting neutrino propagation or detection, including Lorentz-invariance violation~\cite{Satunin:2025uui,Yang:2025kfr,Cattaneo:2025uxk}, exotic matter effects~\cite{Dev:2025czz}, resonance-assisted active–sterile oscillations \cite{Brdar:2025azm}, neutrino self-interactions~\cite{He:2025bex}, and dark matter–neutrino scattering~\cite{Mondol:2025uuw,Esteban:2025wbv,Bertolez-Martinez:2025trs}.  

As emphasized previously in the context of future
neutrino observatories, such as GRAND \cite{GRAND:2018iaj}, POEMMA \cite{POEMMA:2020ykm}, or IceCube-Gen2~\cite{IceCube:2019pna}, in~\cite{Huang:2022pce},
neutrinos offer a unique opportunity to test the transition dipole moment between active and
sterile neutrinos with TeV-scale masses. The detection of KM3NeT-230213A provides the first
real data at these extreme energies and gives us an unexpected opportunity to explore this portal. In this work, we study the
implications of the event for the transition dipole scenario and assess the corresponding
constraints. We prefer to remain agnostic as to the origin of the neutrino and instead focus on local effects. The transition dipole interaction leads to repeated upscattering of the active neutrino into a sterile state, followed by prompt decay back into an active neutrino and a photon. While this process does not remove the neutrino from the beam, it degrades its energy efficiently. As the event is already highly unusual and reconstructed at an extreme energy, a significant energy degradation along the trajectory to the detector is not plausible. We, therefore, constrain the parameters of the dipole portal by requiring that energy losses induced by local interactions do not substantially reduce the neutrino energy
before detection.

This work is organized as follows: In Sec.~\ref{sec:event}, we discuss the KM3-230213A event and summarize the aspects of the observation that are important for our analysis.  Next, in Sec.~\ref{sec: Transition Dipole Portal} we introduce the transition dipole portal and compute the key contribution to the cross section at Earth.  We then move on and estimate the limit on the dipole portal strength $\mu$ based on the optical depth in Sec.~\ref{sec:limits}. Finally, we present our conclusions in Sec.~\ref{sec:conclusions}.

\section{THE EVENT KM3-230213A}
\label{sec:event}

 We will briefly review the properties of the event KM3NeT-230213A that are of interest to us in the following.  The track topology and arrival direction indicate that the event was induced by a muon neutrino undergoing a charged-current interaction outside the instrumented volume, producing a high-energy muon that traversed the detector. 
The probability of a background origin is very low, and a cosmic origin is preferred. Further interpretation remains contingent upon future detections of similar events.

The ultrahigh‑energy neutrino event was reconstructed using the Cherenkov light emitted by the muon traversing the detector volume. The detector's optical modules capture the spatial and temporal distribution of the photons, from which the muon’s trajectory, direction, and light yield profile are reconstructed. The total light yield and the track length are correlated with the muon energy loss in the detector volume, thus providing a lower bound on the muon energy in the detector.
From the reconstructed muon energy, the Collaboration performed an extrapolation to the original production of the muon that accounts for energy losses outside the detector. This leads to a determination of the initial muon energy of $\sim 120$ PeV. 
Taking into account the energy transfer in a charged-current interaction and assuming an origin from an all-sky flux with a spectrum $\propto E^{-2}$, the best estimate for the parent neutrino energy is $220$ PeV, while the $90\%$ containment is placed at $72$ PeV to $2.6$ EeV.
We will take the best estimate as our default choice for the neutrino energy and use the $90\%$ upper and lower limits to estimate the effect of the energy uncertainty on our observables.

The arrival direction of the event is almost horizontal. The best fit gives a zenith angle of approximately $0.6^\circ$ above the horizon and  an azimuth of $259.8^\circ$. The primary contributor to the directional uncertainty in the reconstruction of event KM3-230213A is the limited knowledge of the detector's absolute orientation on Earth. Although the positions of the optical modules are continuously monitored via an acoustic positioning system this system alone does not determine the detector's global orientation. The absolute positioning of acoustic emitters is obtained during dedicated sea campaigns, currently yielding an estimated angular uncertainty of approximately $1^\circ$ around each axis. This uncertainty propagates to a $90\%$ confidence region of $2.2^\circ$ in the reconstructed celestial coordinates of the event, making it the dominant systematic uncertainty.
Simulations of muons with energies between 1 and 1000 PeV indicate that the statistical uncertainty in direction reconstruction is significantly smaller, $\lesssim 0.3^\circ$, and thus negligible compared to the systematic pointing error. Planned upgrades, including the deployment of new acoustic emitters with $<1$ m absolute positioning accuracy, aim to refine the detector alignment. This will allow for a recalibration of the data and an improved source localization in the future. 

\begin{figure}[t]
	\centering
	\includegraphics[width=0.7\textwidth]{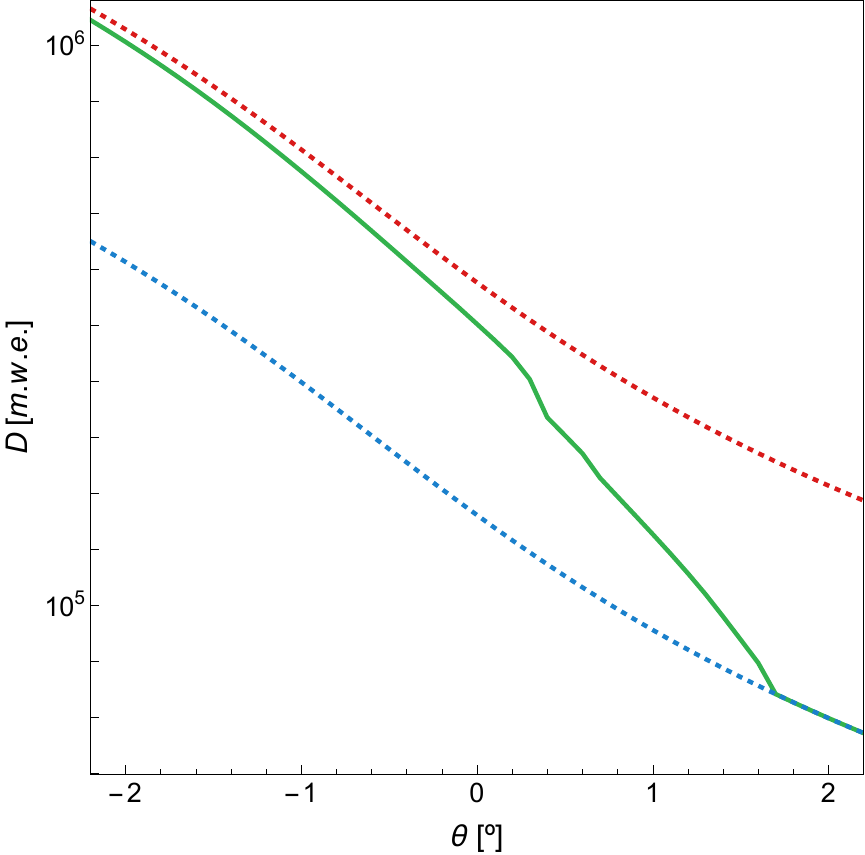}\\
\caption{Depth of the detector $D$ along the neutrino track in units of m.w.e. as a function of the angle with respect to the nominal direction (green solid line). The underwater relief provided by \cite{KM3NeT:2025npi} has been taken and $\rho_{rock}=2.6\rho_{water}$ has been considered. Similar results have been included for comparison purposes, taking into account a medium consisting solely of water (blue dashed line) or rock (red dashed line).} \label{fig:column_density}
\end{figure}

For our purposes, one of the key questions is how much matter the neutrino encounters before it reaches the detector. This is mainly a function of the angle, but, to get an accurate value, we also need to take the local geometry of the Mediterranean seabed into account. We use the underwater relief provided by \cite{KM3NeT:2025npi} and use their suggested value of the density of rock $\rho_{rock}=2.6\rho_{water}$. The depth of the detector $D$ in units of meter water equivalent (m.w.e.) is shown in  Fig. \ref{fig:column_density} as a function of the angle $\theta$ with respect to the nominal arrival direction.  For comparison, the depths for a path through pure water and pure rock are also shown. As can be seen, these nicely bracket the true result. The neutrino starts to encounter rock only for $\theta\lesssim 1.8^\circ$. 
For angles $\lesssim -1^\circ$, the amount of water encountered is small compared to the amount of rock and the line approaches this limit. 
Overall, $D$ varies between $\sim 6.0 \times 10^4 \,\mbox{m.w.e.}$ and $\sim 1.5 \times 10^6 \,\mbox{m.w.e.}$ for angles in the $90\%$ confidence region.

\section{THE TRANSITION DIPOLE PORTAL}\label{sec: Transition Dipole Portal}

Sterile neutrinos can interact with the SM via a transition magnetic dipole moment. The interaction Lagrangian reads 
\begin{align}
    \mathcal{L}_{int}=\mu \bar{\nu}_L \sigma_{\mu \nu} F^{\mu \nu} N + {\rm H.c.}
\end{align}
where $\nu_L$ denotes a left-handed SM neutrino, $F^{\mu \nu}$  is the field strength tensor of electromagnetism,  $N$ is the right-handed neutrino, and  the strength of the interaction is controlled by the parameter $\mu$. The sterile neutrino gets a mass $m_s$ from a Majorana mass term. Here and in the following, we assume that this mass is larger than the active neutrino one. In this case,  the dipole portal allows for upscattering of an SM neutrino into a sterile state and the subsequent decay of a sterile state to a photon and an active neutrino. Depending on the mass and the strength of $\mu$ this leads to a whole range of testable phenomena that can be leveraged to  constrain the strength of the transition dipole moment by lab experiments as well as astrophysical and cosmological observables; see, e.g., \cite{Magill:2018jla,Brdar:2020quo,Chauhan:2024nfa,Miranda:2021kre,Schwetz:2020xra,Bolton:2021pey,Shoemaker:2018vii,Coloma:2017ppo}. Because of the limited energy in these experiments it is relatively easy to constrains small $m_s$. However, LEP and LHC are able to push the limits up to masses of $\sim 100$ GeV and $\sim 1$ TeV, respectively. 
Currently, there are no limits on higher masses.

An active neutrino can upscatter to a heavier sterile neutrino in collisions with SM particles. For high masses, the interaction with electrons does not provide the required center-of-mass energy, and, therefore, one needs to consider neutrino scattering on nuclei. Depending on the momentum transfer to the nucleus the cross sections can be split into three regimes: (i) coherent scattering on the whole nucleus, (ii) diffractive scattering where a nucleon is kicked out of the nucleus, and (iii)  deep inelastic scattering (DIS) where the momentum transfer is large enough to probe the internal structure of the nucleon such that the interaction is best modeled as a scattering on the partonic constituents of the nucleon directly. The elastic and diffractive scattering regimes turn out to be subleading at the large energies we are interested in.  We model them following the method outlined in \cite{Magill:2018jla,Huang:2022pce} and include them in our analysis. To avoid cluttering the manuscript, we relegate their description to the Appendix \ref{Appx:Elastic_inelastic} and focus on the DIS cross section here. 

\begin{figure}[t]
	\centering
	\includegraphics[width=0.7\textwidth]{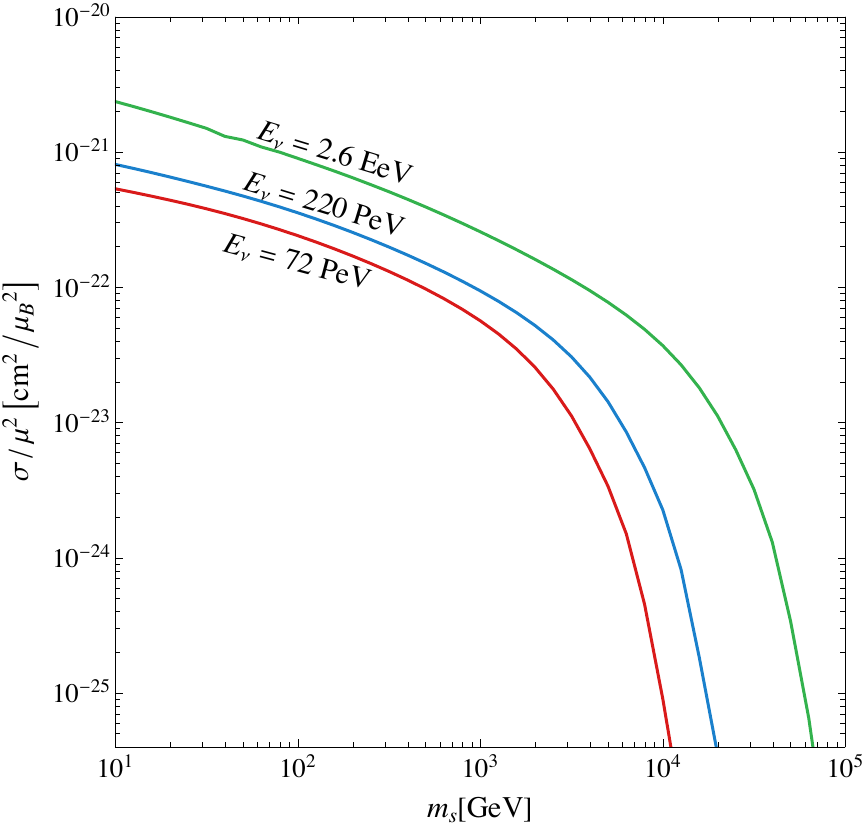}\\
\caption{The nucleon-averaged neutrino cross sections of conversion to a sterile
neutrino per unit of  $\mu$ squared as a function of $m_s$ for the extremal and favored values of the $E_\nu$ at a confidence level of $90\%$. The collision target here is fixed as ${\rm H}_2{\rm O}$ for demonstration. 
} \label{fig:sigma}
\end{figure}

The differential cross section for upscattering in a collision with a massless pointlike fermion is given by
\begin{align}
 \frac{d \sigma}{ d t}   = -\frac{2 \alpha q^2 \mu ^2 \left(m_s^4-m_s^2 (2 s+t)+2 s (s+t)\right)}{s^2 t}
\end{align}
where $s$ and $t$ are Mandelstam variables and $q$ denotes the charge of the fermion.
In order to arrive at the total cross section for scattering on nucleons we need to fold this with the parton distribution functions (PDFs). This leads to 
\begin{align}
    \sigma(E_\nu) = \sum_i\int^1_{x_{min}} dx \int^{t_{1}}_{t_{0}} dt \; \frac{d \sigma}{ d t}(\hat{s}) \; f_i(x,Q)
\end{align}
where $\hat{s}= 2 m_s E_\nu x$ denotes the partonic center-of-mass energy and $x$ gives the fraction of the total nucleon energy carried by the parton. The function $f_i$ is the PDF of species $i$ as a function of $x$ and depends on both $x$ and the momentum transfer $Q=\sqrt{-t}$. The lower limit of integration $x_{min}=(E_\nu-\sqrt{E_\nu^2 -m_s^2})/m_s$ can easily  be found from kinematics, as can the limit $t_{0}= m_s^2-s$. The remaining limit, $t_{1}$, is given by the breakdown of the DIS approximation. We adopt the conventional choice $Q_{min}^2= -t_{1}=4$ GeV$^2$ \cite{Blumlein:2012bf}. We use the {\it Mathematica} implementation of the PDF set \cite{Martin:2009iq}  available at \cite{MSTW2008}. In
Fig.~\ref{fig:sigma} we show $\sigma$ including all contributions as a function of  $m_s$ for three representative values of $E_\nu$ and an H$_2$O target.

\section{LIMITS ON $\mu$}
\label{sec:limits}

The extreme energy of the KM3NeT-230213A event allows us to set new constraints on the
transition dipole moment $\mu$. While the origin of the neutrino is unknown, its trajectory
toward the detector is reasonably well constrained. As discussed in Sec.~\ref{sec:event},
the amount of matter traversed by the neutrino before reaching the detector is substantial.
This makes the event sensitive to new physics effects inducing interactions with matter
along the path.

In the transition dipole portal, an active neutrino can upscatter into a heavier sterile
state, which subsequently decays promptly into a photon and an active neutrino. As a result,
the neutrino is not removed from the beam. Instead, the process leads to a regeneration
of the active neutrino albeit at a a significantly lower energy. For ultrahigh-energy neutrinos,
each upscattering event typically transfers an $50\%$ of the neutrino
energy to the emitted photon, such that the regenerated active neutrino carries
approximately half of the incoming energy.

The cumulative effect of these interactions can be described by an energy transport
equation. In the continuum limit, the evolution of the neutrino energy $E$ along the path
length $l$ is given by
\begin{align}
    \frac{1}{E}\frac{dE}{dl}
    = -\frac{\Delta E}{E}\,\Gamma(E,l)
    \simeq -\frac{1}{2}\,\Gamma(E,l),
\end{align}
where $\Gamma(E,l)=\sigma(E,m_s,\mu,l)n(l)$ is the interaction rate and $n(l)$ denotes the
effective nucleon number density along the trajectory.

If the energy and spatial dependence of the interaction rate factorize,
$\Gamma(E,l)=\sigma(E)\,n(l)$, the transport equation can be rearranged and integrated,
yielding
\begin{align}
    \int_{E_0}^{E_f} -\frac{2}{\sigma(E)E}\,dE
    = \int_0^{L} n(l)dl \equiv \Sigma,
\end{align}
where $E_0$ and $E_f$ denote the neutrino energy at the surface of Earth and at the
detector, respectively, and $\Sigma$ is the known column depth along the neutrino trajectory.

Since KM3NeT-230213A was reconstructed at an extreme energy, it is not plausible that the
neutrino experienced a large number of such energy-degrading interactions before reaching
the detector. We, therefore, conservatively require that the cumulative energy loss along the
trajectory does not exceed an order-ten effect and demand $E_f\geq 0.1 E_0$. This requirement can be used to estimate an upper bound on the transition dipole moment $\mu$ by evaluating the above integral
for a given trajectory and observed final neutrino energy.

\begin{figure}[htbp]
	\centering
\includegraphics[width=0.95\textwidth]{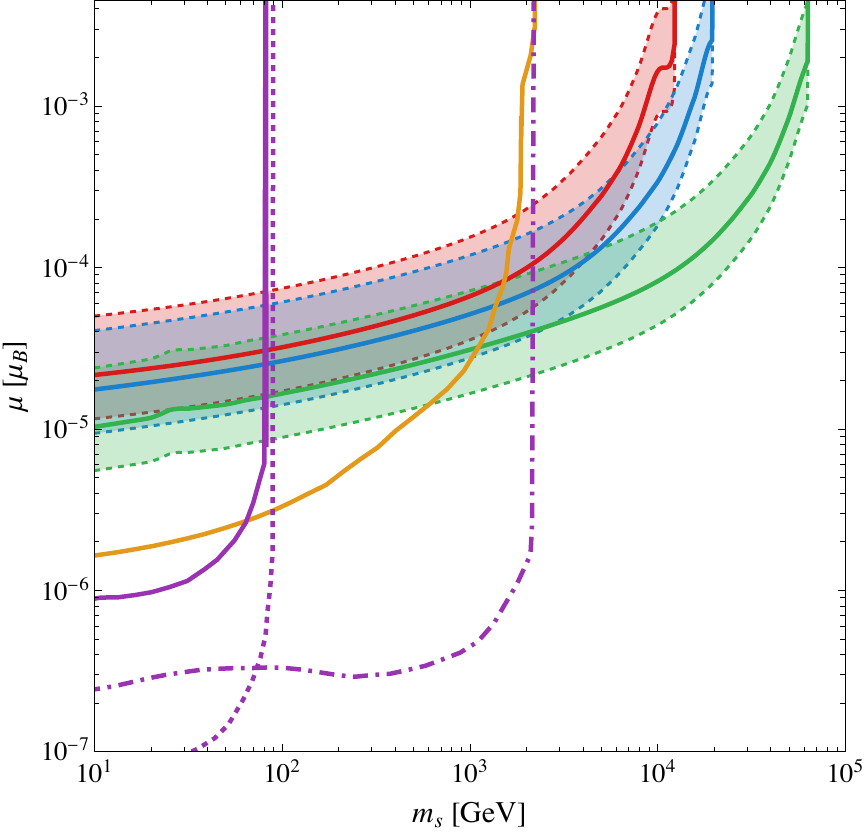}
\caption{Bounds for $\mu$ for the KM3NeT event as a function of $m_s$. Each color band collects the upper limits for different trajectories, i.e., $-2.2^\circ\leq\theta\leq2.2^\circ$ being the angle with respect to the nominal direction (solid line). The red region corresponds to $E_\nu=72$ PeV, the blue region to $E_\nu=220$ PeV, and the green region to $E_\nu=2.6$ EeV. The golden line based on IceCube observations is taken from \cite{Huang:2022pce}; the purple solid, dashed, and dot-dashed line based on LEP ($d_\gamma$ and $d_{\gamma Z}$) and LHC, respectively, are taken from \cite{Magill:2018jla}. }\label{fig:Final_Figure}
\end{figure}

The resulting exclusion contours in the $(m_s,\mu)$ plane are shown in
Fig.~\ref{fig:Final_Figure}. The red,
blue, and green bands correspond to the lower ($72~\mathrm{PeV}$), nominal
($220~\mathrm{PeV}$), and upper ($2.6~\mathrm{EeV}$) estimates of the neutrino
energy, respectively, whose limits are given by the arrival directions within the
$90\%$ confidence region, namely, $\theta=+2.2^\circ$ and
$\theta=-2.2^\circ$ relative to the nominal trajectory. For reference, the golden line indicates the limit \cite{Huang:2022pce} derived based on IceCube
results~\cite{IceCube:2020rnc}, while the purple lines based on LEP and LHC searches are taken from \cite{Magill:2018jla}.
One can see that the strength of the bound depends both on the assumed neutrino energy
and on the trajectory through Earth. As expected, larger path lengths correspond to
a stronger attenuation and, therefore, tighter limits on $\mu$, while shorter trajectories
weaken the constraint. Similarly, increasing the neutrino energy enhances the available
center-of-mass energy in the scattering and improves the sensitivity, which explains why
the green curves ($E_\nu=2.6~\mathrm{EeV}$) yield the most restrictive bounds. In contrast,
for the lower energy estimate ($E_\nu=72~\mathrm{PeV}$, red curves), the constraint is weaker
by more than an order of magnitude. Overall, across all angles and energies, the limits
obtained from KM3NeT-230213A surpass the existing IceCube and collider bounds in the region
of sterile neutrino masses above the TeV scale, thereby extending the reach of neutrino
dipole searches into a previously inaccessible parameter space. This is particularly exciting since future $e^+e^-$ colliders are not expected to probe significantly higher sterile-neutrino masses than present collider bounds in the dipole-portal scenario. Note, however, that a future muon colliders could extend the sensitivity to multi-TeV masses~\cite{Brdar:2025iua}.

We emphasize that the limits derived here are based on a single event and should, therefore,
be interpreted with caution. Nevertheless, they demonstrate the potential of ultrahigh-energy
neutrino observations to explore new physics at energy scales well beyond current collider
reach. Future detections of similar events will allow these limits to be refined and
established on firmer statistical grounds.

\section{SUMMARY}
\label{sec:conclusions}

The measurement of event KM3-230213A opens unique opportunities for the study of physics beyond the SM.
In this work, we investigated its implications for the transition dipole portal to sterile neutrinos. The high energy of the incident neutrino implies that the production of sterile neutrinos with masses in the multi-TeV range becomes kinematically allowed for the first time. The improvement in the mass reach compared to IceCube is roughly one order of magnitude, which can easily be understood from the fact that $s\propto \sqrt{E_\nu}$ and the KM3NeT event was approximately 2 orders of magnitude more energetic than the highest neutrinos seen at IceCube.
This allows us to test new regions  of the parameter space and put new constraints on the 
interaction strength $\mu$.

Since the origin of the neutrino is poorly constrained at the moment, we focus on local effects and study the energy loss caused by interaction with material at Earth. We compute the interaction rate of ultrahigh-energy neutrinos along the path to the detector. This depends both on the new physics parameters that enter into the cross section and on the trajectory to the detector. We compute the cross section taking different contributions from elastic scattering on nuclei, diffractive scattering on nucleons, and deep inelastic scattering into account. The exact trajectory also has a significant impact.
Even though the neutrino event was observed close to the horizon, the amount of water and rock that the neutrino had to traverse on its way to the detector is significant; e.g., for the preferred arrival direction it corresponds to approximately $4 \times 10^5$ meter water equivalent. Note, however, that, due to errors in the determination of the arrival direction, the depth of the effective detector can vary by a factor $\approx 20$ within the $90\%$ confidence level interval. It is expected that this uncertainty can be reduced considerably in the future, since it is dominated by a systematic error in the determination of the overall orientation of the detector, which can be reduced by a dedicated measurement campaign.    

We estimated a limit on $\mu$ as a function of the sterile neutrino mass by requiring that interactions with Earth do not induce a excessive degradation of
the neutrino energy before detection. For $m_s \lesssim 1$  TeV, there are already quite stringent limits from collider searches and IceCube. These are more constraining where they apply. However, they cannot test masses in the multi-TeV regime, which leaves room for the KM3NeT observation to contribute, and we find that our limits are the most constraining there. This is very exciting and demonstrates that the observation of even a single ultrahigh astrophysical neutrino can add to our understanding of physics beyond the standard model. 

\acknowledgments
A.M.O. acknowledges support from the Generalitat Valenciana programs Plan GenT Excellence Program CIDEGENT/2020/020, PROMETEO/2019/083, and CIACIF/2021/260.

\appendix
\begin{section}{APPENDIX: ELASTIC AND DIFFRACTIVE SCATTERING}
\label{Appx:Elastic_inelastic}

In this appendix, we provide ingredients to compute the cross sections for coherent and diffractive  scattering of high-energy  neutrinos into heavy sterile neutrinos by a transition magnetic dipole moment 
 following the presentation in~\cite{Magill:2018jla,Huang:2022pce}.
The DIS contribution is discussed in Sec. \ref{sec: Transition Dipole Portal}.

\subsection{General formalism}

The squared matrix element for neutrino upscattering via virtual photon exchange reads
\begin{equation}
    |\mathcal{M}|^2 = \frac{\mu^2 e^2}{q^4} L^{\mu\nu} W_{\mu\nu},
\end{equation}
where \(\mu\) is the transition dipole moment, \(q^2 = -Q^2\) is the momentum transfer squared, and \(L^{\mu\nu}\) and \(W^{\mu\nu}\) are the leptonic and hadronic tensors, respectively.

The leptonic current from the dipole vertex is
\begin{equation}
    j^\mu = \langle N | \hat{j}^\mu | \nu \rangle = 2\, \bar{u}_\nu(p) P_R \sigma^{\mu\alpha} u_N(p') q_\alpha,
\end{equation}
leading to the spin-summed leptonic tensor
\begin{equation}
    L^{\mu\nu} = 4\, \mathrm{Tr} \left[ \slashed{p} P_R \sigma^{\mu\alpha} q_\alpha (\slashed{p}' + m_s) \sigma^{\nu\beta} q_\beta \right].
\end{equation}
The hadronic tensor can be written in terms of the structure functions \(W_1\) and \(W_2\) as
\begin{equation}
\label{eq:W}
    W^{\mu\nu} = \left(-g^{\mu\nu} + \frac{q^\mu q^\nu}{q^2}\right) W_1(Q^2) + \frac{1}{M^2} \left(k^\mu - \frac{k \cdot q}{q^2} q^\mu \right)\left(k^\nu - \frac{k \cdot q}{q^2} q^\nu \right) W_2(Q^2),
\end{equation}
where \(M\) is the mass of the scattering target, i.e. a nucleus or nucleon in this context, and \(k\) its initial four-momentum.

The hadronic vector current $J_\mu$ for a nucleon or nucleus can be written in the general form
\begin{equation}
    \langle A'|\hat J_\mu|A\rangle 
    = \bar{u}(k') \left[ \gamma_\mu\, F_1(Q^2) 
    + i\, \sigma_{\mu\alpha} \frac{q^\alpha}{2M}\, F_2(Q^2) \right] u(k).
\end{equation}

From this expression, one obtains the corresponding form factors entering the hadronic tensor 
$W_{\mu\nu} = j_\mu^H (j_\nu^H)^\dagger$ in Eq.~(\ref{eq:W}). They can be expressed as
\begin{align}
    W_1(Q^2) &= 2 Q^2\, \big[F_1(Q^2) + F_2(Q^2)\big]^2, 
    \\[6pt]
    W_2(Q^2) &= 8 M^2 \Big[F_1^2(Q^2) 
    + \frac{Q^2}{4M^2}\, F_2^2(Q^2)\Big].
\end{align}

Thus the differential cross section for elastic and coherent scattering can then be written as
   \begin{align}
\frac{d\sigma}{dt} = {} & \frac{2 \alpha F_1^2 \mu^2}{t^2(M^2 - s)^2} \bigg[
 - 2 M^2 m_s^4 - t^2 (2 s - 2 M^2 - m_s^2) - t (\, 2 s^2 - 4 M^2 s - 2 m_s^2 s + 2 M^4 + m_s^4)
\bigg] \\
& + \frac{\alpha F_2^2 \mu^2}{4t (M^3 - M s)^2} \bigg[
4 M^2 m_s^4 - t^2 ( 4 s - m_s^2 )- t^3 - t (4 s^2 - 8 M^2 s - 4 m_s^2 s + 4 M^4 + 4 M^2 m_s^2)
\bigg] \\
& + \frac{2 \alpha F_1 F_2 \mu^2}{t (M^2 - s)^2}\bigg[
 - 2 m_s^4 + t m_s^2 + t^2
\bigg]
\end{align}
where $s$ and $t$ are the usual Mandelstam variables, the momentum transfer $Q$ is related to $t$ via \(t = -Q^2\), \(s\), and \(m_s\) denotes the sterile neutrino mass.

\subsection{Elastic scattering regime}

In the elastic regime (\(Q^2 \lesssim 0.5~\mathrm{GeV}^2\)), the neutrino scatters elastically off the entire nucleus. The cross section is enhanced by the nuclear charge \(Z\), but we need to take the finite size of the nucleus and its internal structure into account. This can be done by introducing nuclear form factors.
We use
\begin{align}
    F_1(Q^2) &= Z F_{\mathrm{WS}}(Q^2), \\
    F_2(Q^2) &\approx 0,
\end{align}
where \(F_{\mathrm{WS}}(Q^2)\) is a simple Woods–Saxon nuclear form factor defined by
\begin{equation}
    F_{\mathrm{WS}}(Q^2) = \frac{3 \pi a}{r_0^2 + \pi^2 a^2} \frac{\pi a \coth(\pi Q a) \sin(Q r_0) - r_0 \cos(Q r_0)}{Q r_0 \sinh(\pi Q a)},
\end{equation}
with \(a = 0.523~\mathrm{fm}\) and \(r_0 = 1.126 A^{1/3}~\mathrm{fm}\), where \(A\) is the nuclear mass number.

\subsection{Diffractive scattering regime}

In the diffractive regime, neutrinos scatter quasielastically on individual nucleons within the nucleus.
Finite-size effects and the internal structure of the nucleons can be incorporated by introducing form factors for the nucleons.
The hadronic current is parametrized as
\begin{equation}
    \langle N' | \hat{J}^\mu | N \rangle = \bar{u}(k') \left[ \gamma^\mu F_1(Q^2) + i \frac{\sigma^{\mu\alpha} q_\alpha}{2 M} F_2(Q^2) \right] u(k),
\end{equation}
yielding structure functions
\begin{align}
    W_1(Q^2) &= \frac{1}{2} \left( Q^2 F_1^2(Q^2) + F_2^2(Q^2) \right), \\
    W_2(Q^2) &= \frac{1}{4 M^2} \left( 8 M^2 F_1^2(Q^2) + Q^2 F_2^2(Q^2) \right).
\end{align}

The nucleon form factors \(F_1^{p,n}\) and \(F_2^{p,n}\) are related to the Sachs charge and magnetic form factors \(G_E^{p,n}\) and \(G_M^{p,n}\) by \cite{Perdrisat:2006hj}
\begin{align}
    G_E^{p,n}(Q^2) &= F_1^{p,n}(Q^2) - \frac{Q^2}{4 M^2} F_2^{p,n}(Q^2), \\
    G_M^{p,n}(Q^2) &= F_1^{p,n}(Q^2) + F_2^{p,n}(Q^2).
\end{align}
Similar to \cite{Magill:2018jla,Huang:2022pce}, we adopt the dipole parametrization  of the form factor which is known to provide a good fit to scattering data \cite{Perdrisat:2006hj}:
\begin{equation}
    G_D(Q^2) = \left(1 + \frac{Q^2}{0.71~\mathrm{GeV}^2} \right)^{-2}, \quad \mu_p = 2.793, \quad \mu_n = -1.913,
\end{equation}
with the assignments
\begin{equation}
    G_E^p = G_D, \quad G_E^n = 0, \quad G_M^{p,n} = \mu_{p,n} G_D.
\end{equation}

\end{section}

\bibliographystyle{JHEP}
\bibliography{paper.bib}

\end{document}